\documentclass[aps,prl,twocolumn,superscriptaddress,showkeys]{revtex4-2}  
\usepackage{graphicx}  
\usepackage{dcolumn}   
\usepackage{bm}        
\usepackage{amssymb}   
\usepackage{epsfig}
\usepackage{caption}
\usepackage{color}

\usepackage{booktabs}
\usepackage{xcolor}
\usepackage{colortbl}

\makeatletter
\renewcommand\@makecaption[2]{%
	\par
	\vskip\abovecaptionskip
	\begingroup
	\small\rmfamily
	\begingroup
	\samepage
	\flushing
	\let\footnote\@footnotemark@gobble
	\@make@capt@title{#1}{#2}\par
	\endgroup
	\endgroup
	\vskip\belowcaptionskip
}
\makeatother

\newif\ifarXiv
\arXivtrue 

\begin{document}
	

	\title{Spontaneous breaking of mirror symmetry beyond critical doping in Pb-Bi2212}

	\author{Saegyeol Jung}
	\affiliation{Center for Correlated Electron Systems, Institute for Basic Science, Seoul, 08826, Korea}
	\affiliation{Department of Physics and Astronomy, Seoul National University, Seoul, 08826, Korea}
	
	\author{Byeongjun Seok}
	\affiliation{Center for Correlated Electron Systems, Institute for Basic Science, Seoul, 08826, Korea}
	\affiliation{Department of Physics and Astronomy, Seoul National University, Seoul, 08826, Korea}
	
	\author{Chang jae Roh}
	\affiliation{Center for Correlated Electron Systems, Institute for Basic Science, Seoul, 08826, Korea}
	\affiliation{Department of Physics and Astronomy, Seoul National University, Seoul, 08826, Korea}
	
	\author{Donghan Kim}
	\affiliation{Center for Correlated Electron Systems, Institute for Basic Science, Seoul, 08826, Korea}
	\affiliation{Department of Physics and Astronomy, Seoul National University, Seoul, 08826, Korea}
	
	\author{Yeonjae Lee}
	\affiliation{Center for Correlated Electron Systems, Institute for Basic Science, Seoul, 08826, Korea}
	\affiliation{Department of Physics and Astronomy, Seoul National University, Seoul, 08826, Korea}
	
	\author{San Kang}
	\affiliation{Center for Correlated Electron Systems, Institute for Basic Science, Seoul, 08826, Korea}
	\affiliation{Department of Physics and Astronomy, Seoul National University, Seoul, 08826, Korea}

	\author{Shigeyuki Ishida}
	\affiliation{National Institute of Advanced Industrial Science and Technology (AIST), Tsukuba, Ibaraki 305-8568, Japan}

	\author{Shik Shin}
	\thanks{Deceased}
	\affiliation{Institute for Solid State Physics (ISSP), The University of Tokyo, Kashiwa, Chiba 277-8581, Japan}

	\author{Hiroshi Eisaki}
	\affiliation{National Institute of Advanced Industrial Science and Technology (AIST), Tsukuba, Ibaraki 305-8568, Japan}
	
	\author{Tae Won Noh}
	\affiliation{Center for Correlated Electron Systems, Institute for Basic Science, Seoul, 08826, Korea}
	\affiliation{Department of Physics and Astronomy, Seoul National University, Seoul, 08826, Korea}

	\author{Dongjoon Song}\email{Corresponding autor: dongjoon.song@ubc.ca}
	\affiliation{Center for Correlated Electron Systems, Institute for Basic Science, Seoul, 08826, Korea}
	\affiliation{Stewart Blusson Quantum Matter Institute, The University of British Columbia, 2355 East Mall $\vert$ Vancouver BC  $\vert$ V6T 1Z4 Canada}

	\author{Changyoung Kim}\email{Corresponding autor: changyoung@snu.ac.kr}
	\affiliation{Center for Correlated Electron Systems, Institute for Basic Science, Seoul, 08826, Korea}
	\affiliation{Department of Physics and Astronomy, Seoul National University, Seoul, 08826, Korea}

	\begin{abstract}
		\vspace{1\baselineskip}
		Identifying ordered phases and their underlying symmetries is the first and most important step toward understanding the mechanism of high-temperature superconductivity; critical behaviors of ordered phases are expected to be correlated with superconductivity. Efforts to find such ordered phases have been focused on symmetry breaking in the pseudogap region while the Fermi liquid-like metal region beyond the so-called critical doping $p_{c}$ has been regarded as a trivial disordered state. Here, we used rotational anisotropy second harmonic generation and uncovered a broken mirror symmetry in the Fermi liquid-like phase in (Bi,Pb)$_{2}$Sr$_{2}$CaCu$_{2}$O$_{8+\delta}$ with $p = 0.205 > p_{c}$. By tracking the temperature evolution of the symmetry-breaking response, we verify an order parameter-like behavior with the onset temperature $T_{up}$ at which the strange metal to Fermi liquid-like-metal crossover takes place. Complementary angle-resolved photoemission study showed that the quasiparticle coherence between $\mathrm{CuO_{2}}$ bilayers is enhanced in proportion to the symmetry-breaking response as a function of temperature, indicating that the change in metallicity and symmetry breaking are linked. These observations contradict the conventional quantum disordered scenario for over-critical-doped cuprates and provide new insight into the nature of the quantum critical point in cuprates.
		
		\vspace{2\baselineskip}
	\end{abstract}

	\maketitle

	\section{1. Introduction}

	A long-standing question in the study of cuprate superconductors is how the normal-state properties are related to the high-transition temperature (high-$T_{c}$) superconductivity~\cite{keimer2015quantum,fradkin2015colloquium}. One of the most exotic normal-state characteristics is a strange metal behavior with linear temperature dependence of the resistivity $\rho\sim T^{n} (n=1)$~\cite{varma2020colloquium}. In the experimentally established doping temperature phase diagram of cuprates (Fig. 1a), the strange metal phase with n = 1 is located above the regions of underdoped pseudogap and overdoped Fermi liquid (FL)-like metal phases with $T$-sublinear ($n<1$ at $T < T^{*}$) and $T$-superlinear ($n>1$ at $T < T_{up}$) resistivities, respectively ~\cite{naqib2003temperature,sterpetti2017comprehensive} (Fig. 1b , Supplementary Fig. 1). The prevailing view is that the $T$-linear resistivity persists down to the zero temperature near the critical doping $p_{c}$ of the underlying pseudogap-FL phase boundary, resulting in a V-shaped strange metallic region from the putative pseudogap quantum critical point (QCP)~\cite{hussey2008phenomenology,cooper2009anomalous}. Therefore, it has been suggested that the strange metal behavior arises from the competition between the quantum and thermal fluctuations, i.e., the so-called “QCP scenario”~\cite{proust2019remarkable}.
	
	In the generic QCP scenario, the under-critical-doped pseudogap and over-critical-doped FL-like metal phases play the roles of the quantum ordered and disordered states, respectively, indicating that the transition from the strange metal to the pseudogap phase at $T^{*}$ is a phase transition with a corresponding symmetry breaking, while that to the FL-like metal phase at $T_{up}$ is a smooth crossover. Indeed, various symmetry breaking across the pseudogap phase boundary has been observed, which is consistent with the QCP scenario~\cite{xia2008polar,sato2017thermodynamic,zhang2018discovery,ishida2020divergent}. On the other hand, the over-critical-doped FL-like metal region has been investigated less systematically, and it is unclear whether it definitively represents a quantum disordered state. In contrast to the long-standing belief, ordered phenomena, such as charge order and ferromagnetism, have been observed in the heavily overdoped region ~\cite{kurashima2018development,peng2018re,miao2021charge}. The discovery of ordered phases is surprising not only because it is incompatible with the QCP scenario, but also because the collective bosonic mode attributed to the order parameter fluctuations can mediate the fermion-fermion interaction, as well as superconductivity. Yet, it still remains an open question whether the FL-like metal phase of cuprates is genuinely related to a symmetry broken state with a hidden order or not.
	
	To detect subtle symmetry breaking, rotational anisotropy second harmonic generation (RA-SHG) has been used to research various quantum materials~\cite{zhao2016evidence,harter2017parity,jin2020observation}. Specifically, as it measures the high-rank (n$>$2) optical susceptibility tensor, it is useful in finding both electronically and magnetically driven symmetry breaking induced by multipolar orders. Indeed, previous RA-SHG studies on undoped Sr$_{2}$CuO$_{2}$Cl$_{2}$ and under-critical-doped YBa$_{2}$Cu$_{3}$O$_{y}$ revealed hidden symmetry breaking inside antiferromagnetic and pseudogap regime, respectively~\cite{zhao2017global,torre2021mirror}. To explore the symmetry evolution in the over-critical-doped region using RA-SHG,  (Bi,Pb)$_{2}$Sr$_{2}$CaCu$_{2}$O$_{8+\delta}$ (Pb-Bi2212) is suitable because its crystallographic structure has higher symmetry compared to other cuprates with monoclinic structures, such as LSCO, YBCO, and pristine Bi2212~\cite{zhao2017global,frison2022crystal}. For example, Fig. 1e shows RA-SHG intensity $I_{SS}^{2\omega}$ of Pb-($T_{c}=79$ K) and pristine Bi2212 ($T_{c}=92$ K) as a function of the  $\varphi$, angle of the incident
	plane rotated around the $c$-axis, measured in the $\mathrm{S_{in}{\text -}S_{out}}$ (SS) geometry at room temperature (Fig. 1d). While mirror symmetry along the crystal axis is broken in pristine Bi2212 due to monoclinic distortion~\cite{kan1992four}, the SHG data of Pb-Bi2212 is quite isotropic, which is beneficial for identifying the emergence of small anisotropy in the SHG response.

	Here, we performed a polarization- and temperature-dependent RA-SHG study, along with complementary angle-resolved photoemission spectroscopy (ARPES) measurements of the over-critical-doped Pb-Bi2212 with $p\sim0.205$ and $T_{up}\sim225$ K. Our SHG investigation revealed the unexpected discovery of mirror symmetry breaking with onset temperature coincident with the strange metal to FL-like metal crossover. Moreover, we found that the symmetry breaking is accompanied by order parameter-like enhancement of not only the SHG response, but also the photoemission quasiparticle coherence with the same transition temperature $T_{up}$. Our observations suggest that the ground state of the FL-like metal in the $p>p_{c}$ region is a quantum ordered state with broken mirror symmetry, which contradicts the conventional QCP scenario.

	\begin{figure}[!h]
		\centering 
		\epsfxsize=8.5 cm \epsfbox{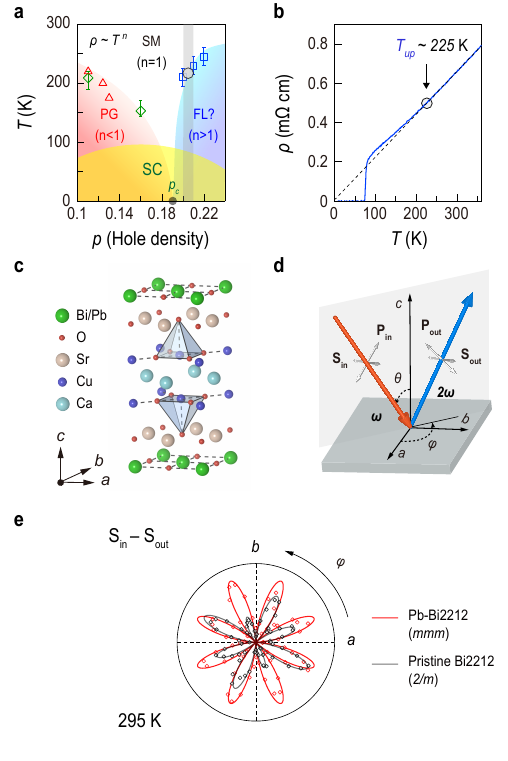}
		
		\caption{\textbf{Orthorhombic crystallographic structure of Pb-Bi2212 with} \boldsymbol{$p>p_{c}$}. \textbf{(a)} Temperature versus doping phase diagram of pristine Bi2212 extracted from the temperature deviation from the linear behavior of in-plane resistivity $\rho_{ab} $(T) (green diamonds~\cite{usui2014doping}, red triangles~\cite{watanabe1997anisotropic}, and blue squares~\cite{kaminski2003crossover}). The doping measured in this study is indicated by the grey line. \textbf{(b)} Temperature-dependent resistivity curve for Pb-Bi2212 with $p\sim0.205$. The grey circle marks $T_{up}$, the onset of upturn deviation from linear resistivity. \textbf{(c)} Crystal structure of Pb-Bi2212. Pb substitution alleviates the structural modulation that exists along the $a$-axis. \textbf{(d)} Schematic illustration of the RA-SHG setup. Linear polarized light, which is parallel (P) or perpendicular (S) to the incidence plane, with frequency $\omega$ focused obliquely onto the (001) surface of the sample. $\varphi$ is the angle of incident plane rotated around the $c$-axis. \textbf{(e)} RA-SHG patterns of Pb-Bi2212 and pristine Bi2212 acquired in $\mathrm{S_{in}{\text -}S_{out}}$ (SS) polarization geometry at room temperature. Data were fitted to the bulk electric quadrupole induced SHG from $mmm$ and $2/m$, respectively.}
		
	\end{figure}

	\begin{figure*}[ht] 
		\centering \includegraphics[width=0.8\textwidth]{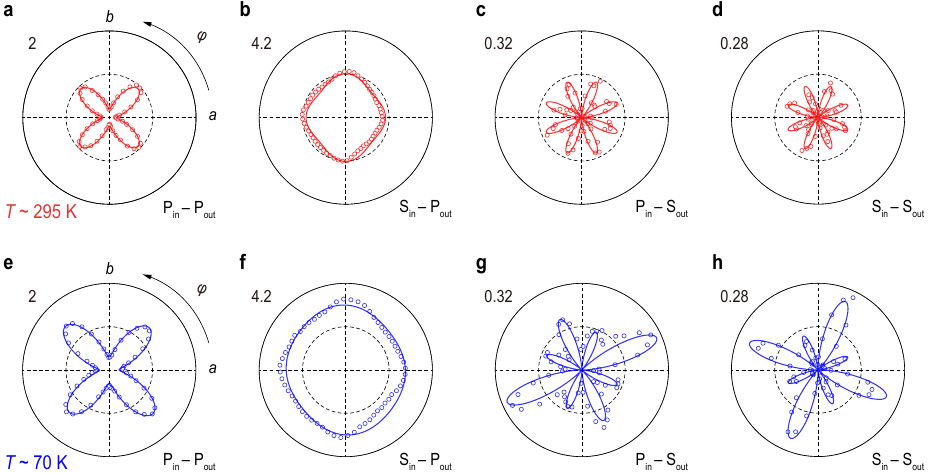}
		\caption{\textbf{Spontaneous mirror symmetry breaking in Pb-Bi2212.}  RA-SHG data of Pb-Bi2212 collected under four polarization geometries: \textbf{(a,e)} PP, \textbf{(b,f)} SP, \textbf{(c,g)} PS, and \textbf{(d,h)} SS at $T\sim295$ K ($T > T_{up}$) (a–d) and $T\sim70$ K ($T < T_{up}$) (e–h). All data sets are presented with the same intensity scale, in which the maximum intensity in the PP geometry at $T\sim295$ K is set to 1. A data set of room and low temperatures for each polarization was obtained from same samples (see Methods for details). The data taken at room temperature are fitted to bulk EQ-induced SHG from $mmm$ (red lines), whereas the low-temperature data are fitted to the coherent superposition of bulk EQ-induced SHG from $mmm$ and $mm'm'$ (blue lines), as described in the main text.}
		
	\end{figure*}

	\section{2. Result}
	Fig. 2 shows experimental and simulated RA-SHG results (open circles and solid lines, respectively) obtained with four polarization geometries at room temperature and near 70 K: $\mathrm{P_{in}{\text -}P_{out}}$ (PP), $\mathrm{P_{in}{\text -}S_{out}}$ (PS), $\mathrm{S_{in}{\text -}P_{out}}$ (SP), and $\mathrm{S_{in}{\text -}S_{out}}$ (SS). As there is no consensus on the point group of Bi-based cuprates, we first determined the crystallographic point group of Pb-Bi2212 at room temperature~\cite{ivanov2018local}. For all of the RA-SHG polar data at room temperature, two mirror symmetries along the crystal axes $a$ and $b$ (two mirror planes of $m_{ac}$ and $m_{bc}$) are clearly identified. We simulated RA patterns from bulk crystallographic point group candidates within these two mirror symmetries and twofold rotational symmetry: centrosymmetric $mmm$ and noncentrosymmetric $mm2$. We found that bulk electric quadrupole (EQ, $Q_{ij}^{2\omega}\sim \chi_{ijkl}^{EQ} E_{k} E_{l})$) SHG derived from the $mmm$ point group $I_{SS}^{2\omega} \propto sin^{2}2\varphi \vert \mathrm{A}~ sin^{2}\varphi+\mathrm{B}~cos^{2}\varphi \vert^{2}$ well fit eight lobes in the SS channel, where A and B are linear combinations of $\chi_{ijkl}^{EQ}$ susceptibility tensor components. Moreover, the fitting results in Figs. 2a–d indicate excellent agreement in the other polarization channels with an EQ SHG response from the centrosymmetric $mmm$ orthorhombic point group. In contrast, we excluded $mm2$ from the point group candidates because the electric dipole (ED, $P_{i}^{2\omega}\sim \chi_{ijk}^{ED} E_{j} E_{k})$ SHG, which is the dominant contribution from noncentrosymmetric $mm2$, is forbidden in the SS polarization channel data (Methods and Supplementary Section 2).
	
	Next, we turned our attention to the temperature-dependent symmetry change. Figs. 2e–h show the RA-SHG at low temperature $T\sim$ 70 K below $T_{up}$ as well as $T_{c}$. Compared to the room temperature results, PP and SP channels show larger SHG intensities while retaining the symmetries. In contrast to PP and SP cases, PS and SS channels exhibit significant amplitude modulation of the lobes across the crystal axes, which is not observed at room temperature. This constitutes clear evidence that the mirror symmetry of the system is broken at low temperature, while the onset temperature of this symmetry breaking remains questionable. We note that the signature of the mirror symmetry breaking is detected only in the polarization channels with S output, at which room temperature data are mostly described by in-plane tensor elements ($\chi_{ijkl}^{EQ}$ elements with $i,j,k,l$ = $x$ or $y$). This implies that the low-temperature SHG response mainly comes from the in-plane symmetry breaking.

	To find the onset temperature of symmetry breaking, we fixed $\varphi$ to the maximum lobe intensity angle and tracked the temperature dependence of the SHG intensity $I^{2\omega}(T)$ in fine temperature steps. We obtained the $\Delta I^{2\omega}(T)$ by subtracting the linear-in temperature background from the $I^{2\omega}(T)$, where the linear-in temperature background is attributed to the linearly decreasing change in the crystallographic $\chi_{ijkl}^{EQ}$ elements~\cite{pavlov1997observation,ron2019dimensional,torre2021mirror} (Supplementary Fig. 2). Figs. 3a–d show $\Delta I^{2\omega}(T)$ for all four geometries normalized to their room temperature values before subtracting the background (Supplementary Fig. 3). With changing temperature, the SHG intensity in all polarization geometries abruptly increases  below $T \sim 225$ K which is close to $T_{up}$ observed in the resistivity data instead of $T_{c}$. This feature is a strong indicator of spontaneous symmetry breaking across the border from a strange metal to FL-like metal phase, in contrast to the long-standing conjecture that $T_{up}$ is associated with a smooth crossover to the disordered state. Moreover, except for the $\mathrm{S_{in}{\text -}P_{out}}$ data, $\Delta I^{2\omega}(T)$ shows obvious order parameter-like behavior below $T_{up}$, indicating that there is an order parameter lowering the symmetry of the system.

	An important open question is whether the $T_{up}$ coincidently meets the onset temperature of the symmetry breaking, or whether they are mutually connected. As $T_{up}$ is an indication of a change in dynamics of charge carriers, the possible emergence of the order parameter is expected to impact the quasiparticle self-energy, which is reflected in the photoemission spectral function. Therefore, we performed complementary ARPES measurements with fine temperature steps to observe the effects of symmetry breaking on the quasiparticles. Figs. 4a–d show ARPES spectra and the second energy derivatives, respectively, above $T_{up}$ at 280 K and below $T_{up}$ at $140$ K, along the Brillouin zone boundary. Compared to the blurred spectral distribution in the $280$ K data, the spectrum at 140 K showed clear bilayer splitting with certain features of the bonding and anti-bonding band. We traced the temperature evolution of the split peaks through the energy distribution curve (EDC) at the anti-nodal point (Fig. 4e, inset). In Fig. 4e, while EDC near room temperature shows a shoulder-like feature around -50 meV, the split peaks become clearly discernible with decreasing temperature, which implies that the doped holes on the two $\mathrm{CuO_{2}}$ bilayers in the unit cell propagate more coherently at lower temperature.
	
	To quantify the quasiparticle coherence and track its temperature evolution, we focus on the temperature dependence of the spectral weight transfer to the coherent split peaks, $W_{coh}(T)$ (Inset of Fig. 4f). In detail, the $\Delta W_{coh}(T)$ is obtained by integrating the spectral weight difference between EDC of 300 K and each temperature for the energy range of -75 meV$ < E{\text -}E_{F} < $0 with subsequent subtraction of the linear-in temperature background, which is considered to be the simple impurity scattering factor (Supplementary Fig. 3). In Fig. 4f, the $\Delta W_{coh}$ is plotted along with $\rho-\rho_{Linear}$ and $\Delta I_{SS}^{2\omega}$, where $\rho_{Linear}$ is the linear component of resistivity obtained by fitting the $\rho(T)$ above $T_{up}$ (dashed line in Fig. 1b); thus, $\rho-\rho_{Linear}$ is a measure of the superlinear component. Remarkably, the $\Delta W_{coh}$ and $\Delta I_{SS}^{2\omega}$ overlap with each other, showing almost the same order parameter-like temperature dependence with the transition temperature $T$ = $T_{up}$, where the superlinear component of resistivity emerges. 
	
	A previous ARPES study proposed an association between the sudden enhancement of quasiparticle coherence
	and the strange metal to FL-like metal crossover, while
	the FL-like metal was still assumed to be disordered~\cite{kaminski2003crossover}. In addition to this long-standing view, our precise quantification of the spectral weight transfer and comparative study with SHG show that the sudden enhancement of coherence is intertwined with the order parameter of the broken symmetry.  Therefore, the results presented
	here suggest that the strange metal to FL-like metal transition at $T_{up}$ in
	Pb-Bi2212 is a phase transition accompanied by mirror symmetry
	breaking.

	\begin{figure}[!h]
		\centering 
		\epsfxsize=8.5 cm \epsfbox{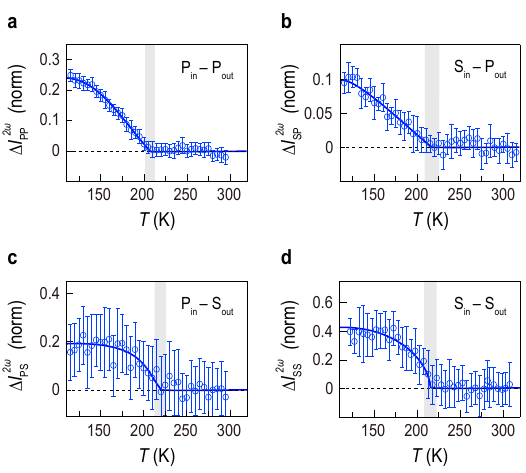}
		
		\caption{\textbf{Temperature dependent RA-SHG pattern amplitude} for \textbf{(a)} PP, \textbf{(b)} SP, \textbf{(c)} PS, and \textbf{(d)} SS in a heating cycle. Data were obtained with $\varphi$ fixed to the maximum SHG lobe at low temperature. Data were first normalized to the values at $T\sim295$ K. Then, a high-temperature linear background was subtracted. The error bars represent the standard deviation of the intensity over 12 independent measurements in the PP, SP, and PS channels. The error bars in the SS channel denote the standard deviation of 60 measurements while heating by 5 K with a ramping rate of 2 K/min (Methods). Blue lines on the data are guides for the eye. The width of the shaded grey interval shows the uncertainty in the transition temperature. }
		
	\end{figure}

	\begin{figure*}[ht] 
		\centering \includegraphics[width=0.85\textwidth]{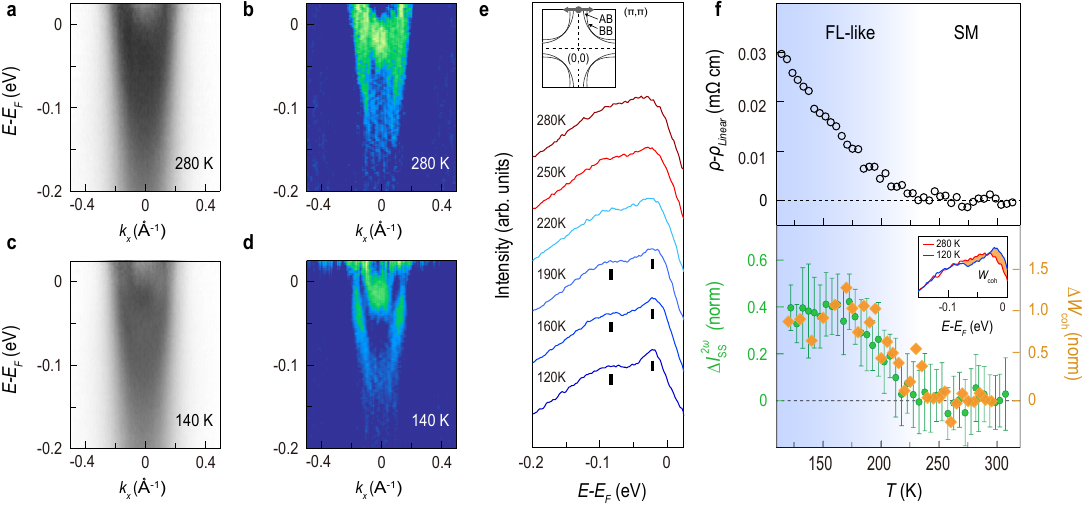}
		\caption{\textbf{Temperature dependence of electronic structure in Pb-Bi2212 across }$\boldsymbol{T_{up}}$.  (a–b) ARPES spectra (left) and second energy derivatives (right) taken above $T_{up}$ at 280 K and (c–d) below $T_{up}$ at 140 K, along the Brillouin zone boundary, as denoted by the solid grey arrow in the inset of (e). All ARPES data were divided by the energy resolution-convoluted Fermi-Dirac distribution function, and the momentum-independent background was then subtracted (see Methods for details). (e) Temperature dependence of EDC at the anti-node (grey dot in the inset of (e)). Each EDC was normalized to the total area over the energy range -200 meV$<E{\text -}E_{F}<$0. Black rectangles denote the bilayer splitting of antibonding band (AB) and bonding band (BB). The inset shows a schematic of the Fermi surface of Pb-Bi2212 composed of the bilayer splitting. (f) $\rho-\rho_{Linear}$  as a function of temperature, where $\rho_{Linear}$  is the linear fitting of $\rho(T)$ between 250 and 300 K (top). Temperature dependence of the integrated spectral weight difference over the energy range -75 meV$<E{\text -}E_{F}<$0 between 300 K and each temperature ($W_{coh}$, orange area in the inset) with subtracting linear-in temperature background, in comparison with the normalized SHG intensity in SS geometry from Fig. 3d (bottom). }
		
	\end{figure*}

	\section{4. Discussion}
	We attempted to narrow down the point group symmetry candidates of the FL-like phase by surveying both crystallographic and magnetic subgroups of the $mmm$ point group, and categorized possible sources of the low-temperature RA-SHG response to the ED contribution from noncentrosymmetric subgroups, and the MD or EQ contribution from centrosymmetric subgroups. Table 1 shows the highest-rank subgroups of the $mmm$ point group and the possible SHG enhancement in each polarization geometry. As susceptibility tensors $\chi_{ijk}^{ED}$, $\chi_{ijk}^{MD}$, and $\chi_{ijkl}^{EQ}$ are invariant under the corresponding symmetry operation of each subgroup, the SHG response can occur in certain polarization geometries (Supplementary Section 2). Therefore, all noncentrosymmetric subgroup candidates can be excluded due to the forbidden ED SHG enhancement in SS geometry, which contradicts our observation in Fig. 3d. As the SHG enhancement at $T_{up}$ occurs in all polarization geometries, we identified the subgroup candidates from Table 1: the EQ process from $2/m$ crystallographic subgroup and EQ process from $mm'm'$ magnetic subgroup, where $m'$ denotes the combination of mirror operation and time reversal.

	We further fitted the low-temperature polar data with the two subgroup candidates of $2/m$ and $mm'm'$, and crosschecked whether the subgroup candidates indeed reflect the symmetry of the system. For the crystallographic subgroup $2/m$, all of the data with four different polarizations are well fitted with $(P_{i}^{2\omega}\sim \chi_{ijkl}^{EQ} \nabla_{j}E_{k} E_{l})$ where $\chi_{ijkl}^{EQ}$ is the nonvanishing susceptibility tensor from $2/m$ (Supplementary Fig. 4). In the case of $mm'm'$, $c$-type contribution $\chi_{ijkl}^{EQ(c)}$ is allowed because time reversal symmetry is broken. Therefore, we fit the data with $P_{i}^{2\omega}\sim (\chi_{ijkl}^{EQ(i)}+\chi_{ijkl}^{EQ(c)}) \nabla_{j}E_{k} E_{l}$, where $\chi_{ijkl}^{EQ(i)}$ is the time-invariant tensor from crystallographic point group $mmm$ and $\chi_{ijkl}^{EQ(c)}$ is the time-noninvariant tensor from the magnetic point group $mm'm'$. The solid lines in Figs. 2e–h show that a coherent superposition of $\chi_{ijkl}^{EQ(i)}$ from $mmm$ and $\chi_{ijkl}^{EQ(c)}$ from $mm'm'$ well reproduces two characteristic features of low-temperature S-output data: the modulation of lobe amplitude, which is a marker of mirror symmetry breaking, and shrinkage along the $a$-axis (Supplementary Fig. 5). These results showed that $2/m$ and $mm'm'$ are suitable candidates to explain both the temperature dependence of SHG intensity and low-temperature polar data.

	Based on these two sub point groups, we discuss the potential physical origins for the order parameter. We first examined the monoclinic structural transition, which enabled the $mmm$ to evolve into the $2/m$ symmetry~\cite{axe1989structural,buchner1994critical}. It is worth noting that, based on our ARPES results, the low-temperature ordering that reduces the symmetry is correlated with enhancement of quasiparticle coherence along the $c$-axis, at least to the distance between the $\mathrm{CuO_{2}}$ bilayers. Therefore, if monoclinic distortion is the origin of the symmetry breaking, it would impact the $c$-axis lattice constant. However, our complementary X-ray diffraction (XRD) study, which focused on the d spacing measurement along the $c$-axis, did not identify any significant anomalies in the temperature evolution of the $c$-axis lattice constant (Supplementary Fig. 4). Moreover, previous XRD and neutron scattering experiments with various Bi-based cuprates found no evidence of such structural transition near $T_{up}$ ~\cite{yang1995thermal,asahi1997thermal}. Nevertheless, we still cannot fully rule out unidentified structural transitions. Therefore, further in-depth study of the structural transition is required.

	Another possibility is that the enhancement of SHG at $T_{up}$ could be attributed to a hitherto unknown bulk order parameter described by the $mm'm'$. Indeed, $mm'm'$ point group incorporates $A_{2g}$ inversion symmetric magnetic order, which is compatible with ferroic order, for example. Intriguingly, several fingerprints of ferromagnetic fluctuations have been reported in the heavily overdoped regime of various cuprates ~\cite{kurashima2018development, sarkar2020ferromagnetic}. In addition, coupling of such magnetic fluctuations with optical phonons is theoretically able to generate the ferroic order with a quasistatic magnetoelectric quadrupole ~\cite{fechner2016quasistatic}. Detection of such multipolar orders via, for example, magnetic neutron scattering and muon spin measurements represents an important direction for further study~\cite{lovesey2019direct}.

	Lastly, our comprehensive study with SHG, ARPES, and resistivity promotes a better understanding of the strange metal phase. We highlight that the $T$-superlinear component of resistivity  $\rho-\rho_{Linear}$ becomes zero as the coherence of the bilayer-split quasiparticle peaks and multipolar order which can be described by the $z$-axis component of the magnetization are significantly suppressed above $T_{up}$ (see Fig. 4). This implies that the decoherence of doped holes between the $\mathrm{CuO_{2}}$ layers and the resulting two-dimensional confinement is an important ingredient of the strange metal behavior in the over-critical-doped regime~\cite{boebinger2009abnormal}.  These results provide experimental support for the extensive theoretical descriptions of non-Fermi liquid behaviors from QCP in two-dimensional metals including cuprates~\cite{stockert1998two,metlitski2010quantum}.

	\definecolor{Gray}{gray}{0.9}

	\begin{figure}
		\hspace*{0cm}
		\setlength{\tabcolsep}{0.25cm}
		\begin{table}
			\footnotesize
			\centering
			\renewcommand{\arraystretch}{1.7}
			\begin{tabular}{c|c|c|c|c|c p{5cm}}
				\hline
				
				\rowcolor[gray]{0.9}[0.25cm]
				\textbf{Point group} & \textbf{Process} & \textbf{PP}  & \textbf{SP}    & \textbf{PS}  & \textbf{SS}  \\
				\hline
				$222$   &  \textit{ED} &  O   & O      & O & --  \\ 
				\hline
				$mm2$   & \textit{ED} & O  & O      & O & --  \\
				\hline
				$2/m$   & \textit{MD} & O   & O      & -- & O  \\ \rowcolor[gray]{0.97}[0.25cm]
				\hline
				$2/m$   & \textit{EQ} & O   & O      & O & O  \\ 
				\hline
				
				$mmm'$  & \textit{ED} & O   & O      & O & --  \\ 
				\hline
				
				$mm'm'$  & \textit{MD} & O   & O      & -- & O  \\ \rowcolor[gray]{0.97}[0.25cm]
				
				\hline
				
				$mm'm'$  & \textit{EQ} & O   & O      & O & O  \\ 
				
				\hline
				
				$m'm'm'$  & \textit{ED} & O   & O      & O & --  \\
				
				\hline

			\end{tabular}
			\caption{Summary of polarization geometry dependent SHG for the highest-rank subgroups of the $mmm$ point group. SHG process in light grey background indicate possible candidates which are consistent with both the temperature dependence of SHG intensity and low temperature polar data. See Supplementary section 2 for detailed equations.}
			\label{Table 1}
		\end{table}
	\end{figure}
	
	\section{Acknowledgements}
	We are grateful to  Y. S. Kim, R. Noguchi, S. S. Huh, H. Y. Choi and A. Hallas for their helpful discussions and useful comments. We appreiciate the technical support on fitting process from B.T. Fichera. This work was conducted under the ISSP-CCES Collaborative Programme and was supported by the Institute for Basic Science in Republic of Korea (Grant Numbers IBS-R009-G2 and Grant No. IBSR009-D1
	). This work was supported by the National Research Foundation of Korea(NRF) grant funded by the Korea government(MSIT) (No. 2022R1A3B1077234) and the JSPS KAKENHI (No. JP19H0582*).
	
	\bibliographystyle{naturemag}
	\bibliography{ref}
	
	\section{Methods}
	
	\section{Material growth}

	Single crystals of $\mathrm{Bi_{1.7}Pb_{0.5}Sr_{2}CaCu_{2}O_{8+\delta}}$ were grown by the floating-zone (FZ) method. The $\mathrm{Pb}$ substitution suppresses the monoclinic superstructure modulation which comes from from the $\mathrm{BiO_{2}}$ layer. The $T_{c}$ was defined as the onset temperature of the Meissner effect, which was measured using the DC magnetic susceptibility module of a Physical Property Measurement System from Quantum Design. Hole doping $p$ was determined via the empirical parabolic relation. $T_{c}$ = $T_{c,max}[1-82.6(p-0.16)^{2} ]$, where $T_{c,max}$ = 95 K for the batch of crystals studied here.
	The crystals used in our experiments were aligned with high accuracy by X-ray Laue diffraction. $a$-axis, which is the $\mathrm{BiO_{2}}$ modulation direction, was assigned by measuring the suppressed superstructure replica of the main electronic band with ARPES (Supplementary Fig. 6).
	
	\section{RA-SHG measurements}

	The RA-SHG measurements were obtained in a rotating incident plane setup based on Fresnel rhomb prism (FR600QM; Thorlabs)~\cite{kim2021compact}. The light source was a Ti:Sapphire laser operating with a repetition rate of 250 kHz, pulse width of 150 fs, and center wavelength of 840 nm (RegA 9000; Coherent). The incidence angle was fixed by specification of Fresnel rhomb prism to $\theta\sim 23^{\circ}$. The incident light was focused to a spot size of $\sim$ 30$\mu$m, with a fluence of $\sim$ 2 mJ/$cm^{2}$ maintained to avoid laser-induced heating effects on the sample only except the polar data in Figs. 2 d and h (Supplementary Fig. 7). SS geometry polar data in Figs. 2d and h were obtained with a fluence of $\sim$ 3 mJ/$cm^{2}$. Each data set at 295 K and 70 K  in Fig. 2 was obtained from same samples, cleaved in air before measurement and pumped  down to the pressure below $2\times 10^{-6}$ Torr  . Additionally, each data set of different polarization geometry was obtained from different samples.  Reflected SHG light was collected using a photomultiplier tube with a lock-in amplifier synchronized to the 2-kHz pulsed output, with a time constant of 3 s. Further details of the experimental setup are described in Ref~\cite{kim2021compact}. We carefully aligned the setup to retain the same incident angle while rotating the incident plane, and confirmed the parallelity of incident polarization and output polarization with a gold mirror and GaAs (110), respectively (Supplementary Fig. 8). When measuring the temperature dependence of SHG with SS geometry in Fig. 3d, we continuously measured the SHG intensity while increasing the temperature with a ramping rate of 2 K/min to overcome the weakest intensity in SS geometry. For the other polarization geometries, PP, SP, and PS, we measured SHG intensity after stabilizing each temperature. The data in this article were not only reproduced by scanning along the sample surface, but also across multiple single crystals grown in different batches (Supplementary Fig. 9).

	\section{Fitting procedure.}
	
	For fitting the RA-SHG data, we used the open source software ShgPy provided by Brian Fichera~\cite{shgpy}. The high- ($T > T_{up}$) and low- ($T < T_{up}$) temperature SHG data were fitted to the expression, $I^{2\omega}\sim \vert \chi_{ijkl}^{EQ(i)} \nabla_{j}E_{k} E_{l}\vert^{2}$, $I^{2\omega}\sim \vert (\chi_{ijkl}^{EQ(i)}+\chi_{ijkl}^{EQ(c)}) \nabla_{j}E_{k} E_{l}\vert^{2}$, where $\chi_{ijkl}^{EQ(i)}$ is the time-invariant tensor from crystallographic point group $mmm$, and $\chi_{ijkl}^{EQ(c)}$ is the time-noninvariant tensor from magnetic point group $mm'm'$. RA-SHG data in all polarization geometries (PP, SP, PS, SS) were fitted simultaneously with 15 nonzero independent tensor elements from $mmm$ ($\chi_{xxxx}$, $\chi_{xxyy}$, $\chi_{xxzz}$, $\chi_{xyyx}$, $\chi_{xzzx}$, $\chi_{yxyx}$, $\chi_{yyxx}$, $\chi_{yyyy}$, $\chi_{yyzz}$, $\chi_{yzzy}$, $\chi_{zxzx}$, $\chi_{zyzy}$, $\chi_{zzxx}$, $\chi_{zzyy}$, and $\chi_{zzzz}$) and 13 nonzero independent tensor elements from $mm'm'$
	($\chi_{xxyx}$, $\chi_{xyxx}$, $\chi_{xyyy}$, $\chi_{xyzz}$, $\chi_{xzzy}$, $\chi_{yxxx}$, $\chi_{yxyy}$, $\chi_{yxzz}$, $\chi_{yyyx}$, $\chi_{yzzx}$, $\chi_{zxzy}$, $\chi_{zyzx}$, and $\chi_{zzyx}$). Here, we assumed that all tensor components were real to reduce the complexity of the computation. The explicit fitting functions are summarized in Supplementary Section 2. Note that the fitting parameter values are not unique solutions of the fitting equations. Therefore, the fitting results only indicate whether a specific point group can reproduce four RA-SHG polar data set. 
	
	\section{ARPES measurements.}

	ARPES measurements were performed with a lab-based system equipped with a He-$I\alpha$ discharge lamp ($hv$=21.2 eV) and Scienta DA30 electron analyzer. The energy resolution of the system was set to 12 meV . For the data used in Fig. 4f, the sample was cleaved at 300 K with a chamber pressure around $9 \times 10^{-11}$ Torr and measured upon cooling. For the data presented in Fig. 4, the ARPES spectra were divided by the energy resolution-convoluted Fermi-Dirac distribution function for each temperature, and a momentum-independent background was obtained by integrating the featureless momentum range -0.66 $\mathrm{\AA^{-1}}$$<k_{y}<$-0.6 $\mathrm{\AA^{-1}}$ was subtracted. The EDCs in Fig. 4e  were obtained in different experiment from the data in Fig. 4f. ARPES data in Fig. 4e were acquired with longer measurement time and larger temperature steps for a high S/N ratio.

\end{document}